\documentclass[sn-mathphys-num]{sn-jnl}
\usepackage{graphicx}%
\usepackage{multirow}%
\usepackage{amsmath,amssymb,amsfonts}%
\usepackage{amsthm}%
\usepackage{wasysym}%
\usepackage{mathrsfs}%
\usepackage[title]{appendix}%
\usepackage{xcolor}%
\usepackage{geometry}
\newgeometry{vmargin={25mm}, hmargin={26mm,26mm}, columnsep={10mm}}

\usepackage{lineno}

\begin{document}


\title[Article Title]{ 
Characteristics of Ge-doped Multi-Mode Fibers in Total Ionizing Dose   }

\author[1]{ \fnm{Datao}   \sur{Gong}} 
\author*[2]{\fnm{Suen}    \sur{Hou}}             
\author[2]{ \fnm{Bo-Jing} \sur{Juang}}
\author[3]{ \fnm{Bin}     \sur{Lin}} 
\author[1]{ \fnm{Chonghan} \sur{Liu}} 
\author[1]{ \fnm{Tiankuan} \sur{Liu}} 
\author[4]{ \fnm{Ming}     \sur{Qi}} 
\author[2]{ \fnm{Yi}       \sur{Yang}} 
\author[1]{ \fnm{Jingbo}   \sur{Ye}}
\author[4]{ \fnm{Lei}      \sur{Zhang}} 
\author[1]{ \fnm{Li}       \sur{Zhang}}
\author[3]{ \fnm{HuiPing}  \sur{Zhu}}

\affil[1]{\orgdiv{Physics}, \orgname{Southern Methodist University},
\orgaddress{\city{Dallas}, \postcode{75205}, \state{TX}, \country{U.S.A}}}

\affil*[2]{\orgdiv{Institute of Physics}, \orgname{Academia Sinica},
\orgaddress{\city{Taipei},  \state{Taiwan}, \postcode{11529}}}

\affil[3]{\orgname{Institute of Nuclear Research},
\orgaddress{\city{Taoyuan},  \state{Taiwan}, \postcode{32545}}}

\affil[4]{\orgdiv{Physics}, \orgname{Nanjing University}, 
\orgaddress{\city{Nanjing}, \state{Jiangsu}, \postcode{210093}, \country{China}}}

\abstract{
\textbf{Purpose:} 
The fiber optical links in 850 nm band with Ge-doped multi-mode (MM) fibers 
are well developed for data transmission at 10 Gbps and higher. 
The applications in nuclear environments require radiation resistance.
The characteristics of Ge-doped MM fibers 
are investigated for Radiation Induced Attenuation (RIA) in Total Ionizing Dose (TID).
\\

\textbf{Methods:}
Commercial samples of Ge-doped MM fibers were
irradiated in Go-60 gamma rays at dose rates of 5  
to 1.4k~Gy(SiO$_2$)/hr. The fiber samples were packaged in
water tanks maintained at constant temperatures in the range of -15 to 45~$^\circ$C. 
The optical power
transmitted through the fibers 
were recorded in irradiation,
and in annealing when the source was shielded.
The measurements of RIA in time are 
analyzed for dose rate and temperature dependences.
\\

\textbf{Results:} 
Ge-doped fiber samples of OM2 to OM4 grades were investigated for attenuation 
of optical power in radiation ionizing dose.
Depending on the fabrication technology, two of the fiber types
show radiation resistance with the RIAs of 0.2~dB/m and 0.05~dB/m, respectively,
for the TID of 300~kGy(SiO$_2$). 
At low dose rate of 5~Gy/hr, the RIA increases steadily and the
annealing of low density ionizing defects does not cause notable deviation.
At 1.4~kGy/hr the accumulated defects result to twice higher RIA 
during irradiation, and is worsen to a factor three in cold temperature.  
However, once the source is shielded the recovery is effective
in a few hours.
\\

\textbf{Conclusion:}
The telecom products of 850 nm Ge-doped MM fibers 
provide high speed communication 
in distances of a few hundred meters. 
The industrial fabrication methods provide fibers 
that can endure radiation ionizing dose for applications in nuclear instrumentation.

}





\maketitle

 \twocolumn

\section{Introduction }

The  fiber optics in 850 nm band 
is well developed for data transmission at 10~Gbps and higher.
The applications in nuclear instrumentation 
can provide high-speed data transmission
in low-mass fibers for distances of a few hundred meters. 
The opto-electronics of
Vertical-Cavity Surface-Emitting Laser (VCSEL) and photodiode (PD)
are durable in radiation of ionizing dose.
The degradation of laser light power,  PD current
and signal noises, due to Non-Ionizing Energy Loss (NIEL)
\cite{OptoNIEL-2007,OptoNIEL-2011,CERN-2011,Johnston-2013,PD-2019},
have been tested to 1$\times 10^{15}$ (1~MeV)~n$_{eq}$/cm$^2$
for  the   high radiation field at the Large
Hadron Collider (LHC).

The transceiver ASICs of laser drivers and photodiode TIAs 
are customized to sustain radiation induced defects~\cite{GBT,VTRX,VTRXplus}.
For example,  a laser driver ASIC have been tested with ionizing dose
of up to 300~kGy, to assure service at LHC~\cite{King2010,MTX-TID2024}.
High-speed radiation tolerant transceivers are developed, which include
a 25~Gbps transmitter~\cite{MTX+}, a 40~Gbps multi-channel module~\cite{QTRX},
and a PAM4 prototype~\cite{PAM4,GBT20}.

Radiation resistant fibers are required  
for nuclear instrumentation.
The telecom-grade Ge-doped fibers have being studied 
for  radiation induced attenuation (RIA) 
\cite{Sensors2024,Girard2019,Berghmans2007},  with
\begin{equation}
 RIA = \frac{IL(t)-IL(t=0)}{Length}
\end{equation}
where $t$ is the total ionizing dose (TID),
and $IL$ the light insertion loss of
\begin{equation}
 IL = 10\log_{10}(\frac{P_T}{P_R}),
\end{equation}
with $P_T$ the optical power transmitted and $P_R$ received.

The fibers investigated for applications at LHC
include a P-doped and a Ge-doped types \cite{Oxford}.
The industrial products of 10 Gbps fibers are mostly 
Ge-doped.
However, depending on the dopants and fabrication 
technologies\footnote{the known mathods are
MCVD, modified chemical vapor deposition; 
and PCVD, plasma chemical vapor deposition, etc.},
the characteristics in radiation of ionizing dose are rather different.  
In the following we report radiation tests with Co-60 
gamma rays for four types of  Ge-doped multi-mode (MM) fibers
acquired from several manufacturers.
The types are labeled as Type-B, M, N, O, respectively,
in OM2 to OM4 grades of 50/125 $\mu$m fiber cores.

The fiber irradiation tests were conducted at the 
Institute of Nuclear Energy Research (INER) Co-60 gamma-ray facility.
In Section~\ref{sec:setup} the preparation of fiber samples
and the test setup  are described. 
Fiber samples sealed in water tanks at constant temperatures
were connected to laser light sources. The transmitted light were
measured continuously.

The radiation induced defects generated in fibers 
were studied for
dose rate and temperature dependences, in the range of
5 to 1.4k~Gy/hr, and the temperatures of -15 to $45\;^\circ$C.
The fiber types showing poor radiation characteristics
are discussed in Section~\ref{sec:nonHard}.
Two of the fiber types show resistance to ionizing dose.
The attenuation of optical power in radiation
and annealing after the source being shielded,
are discussed in Section~\ref{sec:Hard}. 

The Ge-doped MM fibers of telecom grades are practical
choices for high speed data transmission in radiation environments.
Fiber of chosen types can endure ionizing dose. 
A summary on the dose rate and temperature for applications 
is discussed in Section~\ref{sec:summary}.

\section{Co-60 fiber irradiation setup }  
\label{sec:setup}

The radiation studies of Ge-doped fibers were conducted at the INER.
The gamma-ray facility has a large assembly of Co-60 pellets ($\diameter 10$~mm) 
encased and paved into an array of 45$\times$300~cm$^2$.
The Co-60 array is stored in a deep pool filled with demineralized water.
In supervised working hours, it is lifted inside a shielded compartment. 
A conveyor belt circulates cargo to be irradiated.

The fiber irradiation requires long period at stable dose rate and temperature.
The test setup has the fiber samples packaged in water tanks
at fixed positions inside the radiation compartment.
The fiber samples were connected by 40~m patch cords
to external data acquisition system.  The water tanks
were chilled by an external bath or a fridge compressor. 
Thermal couples were attached with the temperatures monitored.

\begin{figure}[bh!] 
  \vspace{2mm}
  \centering
    \includegraphics[width=.88\linewidth]{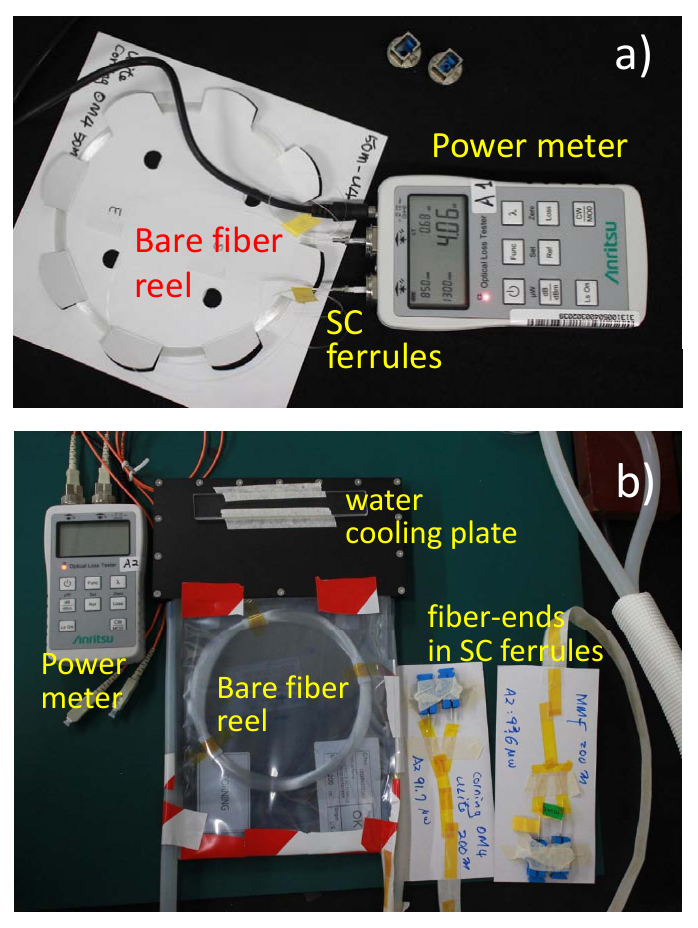}
    \vspace{4mm}
    \caption{ The fiber samples prepared for Co-60 tests are shown for
    a) a bare-fiber reel terminated with SC ferrules which are connected 
    to an  optical loss tester (CMA5, Anritsu);  
    and b) a fiber reel sealed and attached to a water-cooling plate 
    with the fiber ends taped out to SC-SC adapters.   
    In irradiation this package is inserted in a water tank with circulating
    water pumped from an external bath for temperature control.
    \label{fig:sample} }
\end{figure} 

\begin{figure*}[t!] 
  \centering
    \includegraphics[width=.85\linewidth]{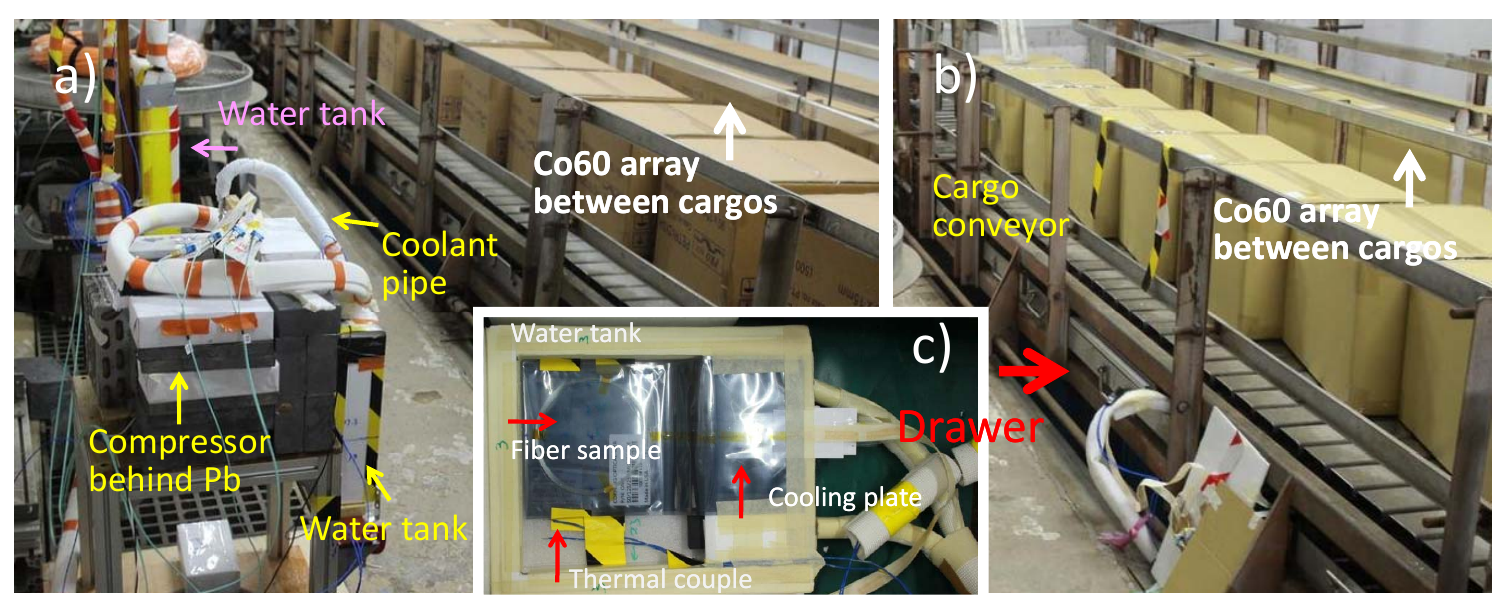}
    \vspace{2mm}
    \caption{The locations of fiber reels in water tanks were chosen
    for dose rates. In supervised working hours,
    the Co-60 source array is lifted from a water pool to the position 
    between the cargo conveyors. Two of the water tanks 
    are shown in a).
    One of them has a compressor evaporator plate inserted for cooling.
    The highest dose rate is located in a drawer
    under the conveyor shown in b). The water tank package in the drawer is
    shown in c). The fiber samples were connected
    by 40~m patch cords to the data acquisition setup outside the radiation area.
  \label{fig:co60tank} }
\end{figure*} 

\subsection{Fiber sample preparation    
and RIA measurement}

Samples of Ge-doped MM fibers were acquired
in grades of OM2 to OM4,
from different vendors and production batches, 
for comparison of uniformity in tests.
The samples were prepared in bare-fiber reels of 4 
to 1k~meters in length. The fiber ends were terminated with
SC type ferrules for connection to the 
laser light sources and power meters.
Depending on the dose rates in irradiation, 
the length of fibers were chosen for the optical power 
attenuation larger than 20~\%,
to be distinguished from systematic fluctuation.

Illustrated in Fig.~\ref{fig:sample}.a is a bare-fiber sample 
connected to a optical power tester. 
In Fig.~\ref{fig:sample}.b the fiber reel is sealed 
with fiber ends ($\sim$50~cm) taped out to SC adapters.
The fiber sample is attached to a water-cooling plate.
In irradiation test, this package would be
inserted in a water tank 
with circulating water pumped from a water-bath 
at a constant temperature of $\pm 0.5\;^\circ$C.

For temperature below $0\;^\circ$C, the fiber samples were
chilled by a fridge compressor, with the evaporator plate
inserted in the water tank. 
In Fig.~\ref{fig:co60tank}.a 
the compressor is shielded inside lead bricks.
The compressor has a thermostat control switch
(XH-W3002), which is fragile to radiation.

The locations of fiber samples in water tanks 
were chosen for the dose rates, by the distance to the Co-60 array 
which is between the cargo conveyors.
A drawer below the front conveyor (Fig.~\ref{fig:co60tank}.b)  
provides a closer position with the dose rate 
reaching 1.4~kGy(SiO$_2$)/hr. 
Shown in Fig~\ref{fig:co60tank}.c is a packaged water tank containing 
two fiber samples in the drawer.

\begin{figure*}[t!] 
  \centering
    \includegraphics[width=.88\linewidth]{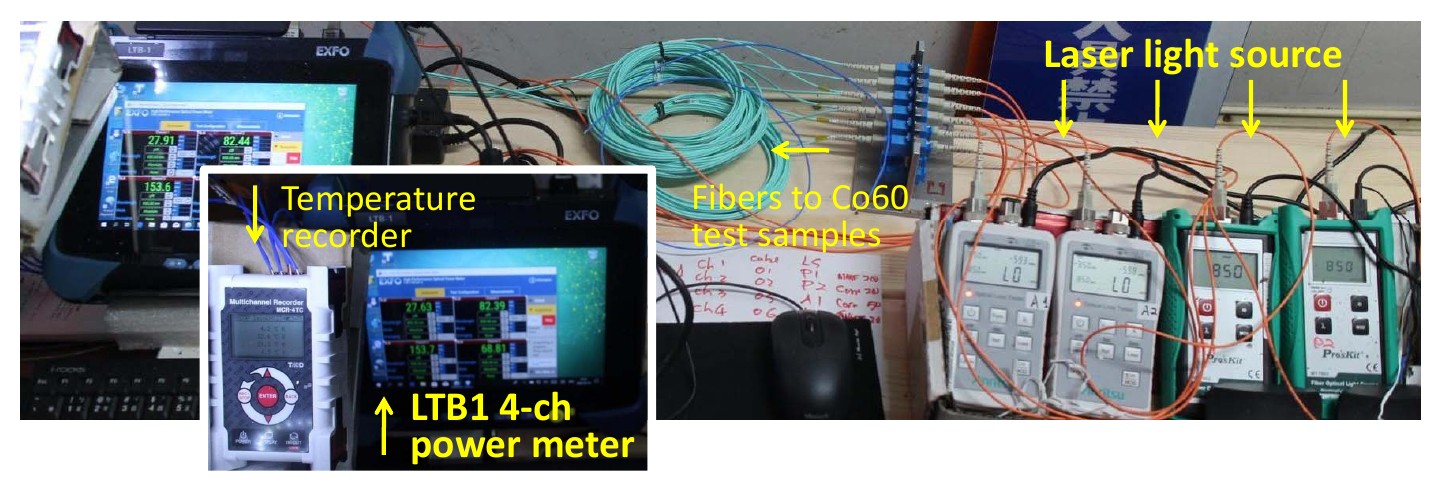}
    \vspace{2mm}
    \caption{The apparatus setup is shown for fiber  
    irradiation with Co-60 gamma rays. 
    On the right are four 850 nm laser light sources 
    (Anritsu CMA5, Pro'sKit MT-7802) connected to 40~m patch cords 
    (blue) to bare-fiber samples in water tanks inside
    the radiation compartment.
    The optical powers returned are recorded by a LTB1 (EXFO),
    and the thermal couples in water tanks by a temperature recorder (MCR-4TC, T\&D). 
    The data taking was conducted every minutes for samples in test setup.
  \label{fig:daq} }
\vspace{5mm}
\end{figure*} 


The picture in Fig.~\ref{fig:daq} shows the data acquisition setup 
outside the radiation compartment. 
The test setup could have four samples irradiated in parallel. 
The 850~nm laser sources were connected to the patch cords 
to fiber samples.
The optical powers returned  
were measured by a LTB1 power meter. 
The thermal couples attached on the samples were
recorded by a temperature recorder.
Once the fiber samples were sealed in water tanks and connected
to laser light sources for optical readout,
the data acquisition proceeded every minute and lasted 
for a few weeks for the full course of a test setup.

\subsection{Dose rate calibration} 

The dose rates on fiber samples were calibrated with 
Alanine pellets (AWM230, Weiser) attached in irradiation.
The total doses received were measured by
an electron paramagnetic resonance (EPR) analyzer 
(Bruker EMS-104) with a precision of better than 1~\%.  

Although the total doses of fiber samples were precisely measured,
the cargos between the fibers and Co-60 source had shielding factors of 
20~\% to 50~\% depending on the cargo densities. 
The dose rates in time were corrected accordingly.
The systematic error on the daily accumulated TIDs is estimated 
to be 10~\%.

The dose measurements with Alanine shall be converted for the 
material of optical fiber, which is fused silica (amorphous SiO$_2$).
The Co-60 decay photons have energies of 1.17~MeV and 1.33~MeV. 
The mass-energy absorption coefficients 
are approximately equal for Alanine and SiO$_2$.
The dose conversion factor applied is 1~Gy(SiO$_2$) = 0.93~Gy(Alanine) 
according to the measurement in \cite{Ravotti}.
In the following, the dose measurements are presented for SiO$_2$.

\section{ Non-radhard  fibers  }  
\label{sec:nonHard}

\begin{figure}[t!] 
  \centering
    \includegraphics[width=.92\linewidth]
    {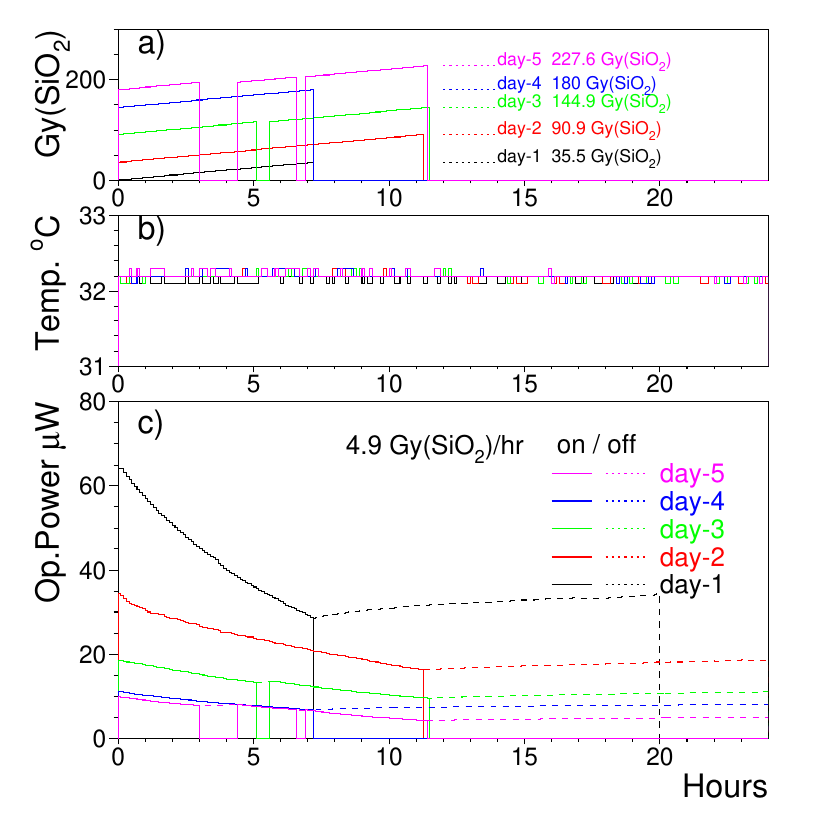}
    \vspace{-2mm}
    \caption{A Type-B Ge-doped OM2 MM fiber sample (10~m) 
    was irradiated in Co-60 gamma rays at a dose rate of
    4.9~Gy(SiO$_2$)/hr, at 32~$^\circ$C.
    The distributions recorded in the initial five days are plotted for 
    a) the accumulated TIDs,
    b) the water tank temperatures, and c) the optical powers transmitted through
    the fiber in irradiation (solid lines) and in annealing with
    the source shielded (dashed lines).
  \label{fig:bsf5gy32oc} }
  \centering
    \includegraphics[width=.92\linewidth]
    {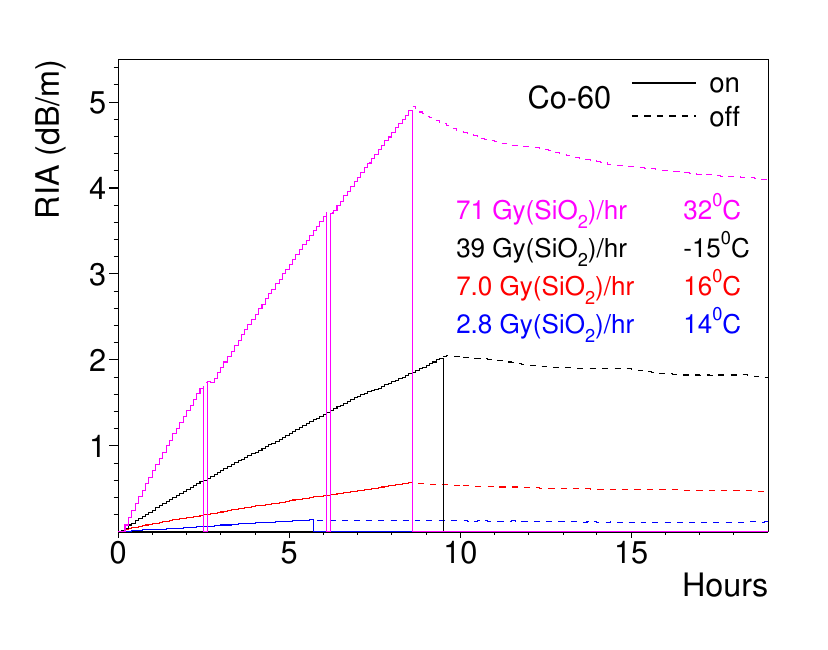}
    \vspace{-2mm}
    \caption{Four of the Type-O Ge-doped OM4 MM fiber samples (4~m)
    were irradiated in Co-60 gamma rays at four dose rates and temperatures.
    The optical power measurements of the first days are plotted
    for the RIAs during irradiation (solid lines) and in annealing
    (dashed lines).
    The RIA reached 4~dB/m at 600~Gy(SiO$_2$).
  \label{fig:ann-ofs} }
\end{figure} 

The tolerance of Ge-doped fibers in ionizing dose can be rather different
depending on the fiber fabrication technologies.
The non-radhard fibers are seen with fast 
optical power attenuation in radiation. 
Plotted in Fig.~\ref{fig:bsf5gy32oc} are the optical power measurements 
of a Type-B OM2 (10~m) sample
irradiated at a low dose rate of 5.9~Gy(SiO$_2$)/hr, 
with the temperature at 32 $^\circ$C.
The measurements in time are presented
for the initial five days.  
The accumulated TIDs and the temperature are plotted in 
Fig.~\ref{fig:bsf5gy32oc}.a and b, respectively.
The optical power dropped faster in the first day (black line).
The annealing after the source being shielded had little recovery. 
With a total of 200~Gy(SiO$_2$), the optical power reduced 
to about 10\% of the original.

The large optical power loss is also observed for the Type-O OM3 fiber.
In Fig.~\ref{fig:ann-ofs} the measurements 
of four samples in the first day are plotted for 
various configurations of dose rates and temperatures.
At the higher dose-rate of 71~Gy(SiO$_2$)/hr, the 
RIA had increased to 4~dB/m
with the accumulated total dose of 600~Gy(SiO$_2$),
after the annealing of about 20 \% recovery.

\begin{figure}[t!] 
  \centering
    \includegraphics[width=.92\linewidth]{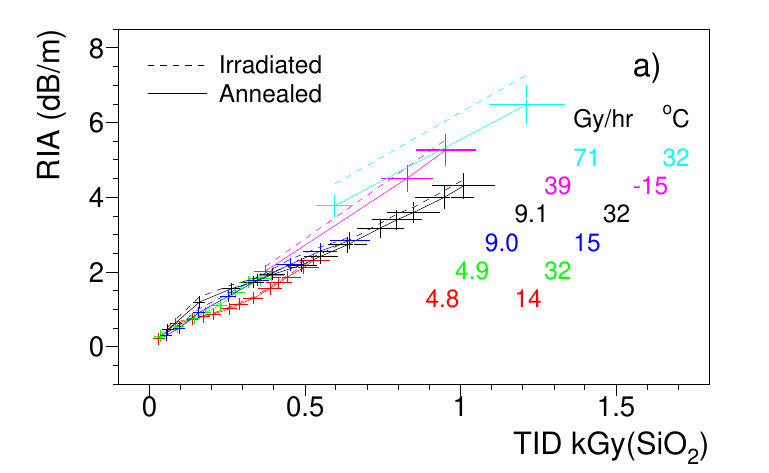}
    
    \includegraphics[width=.92\linewidth]{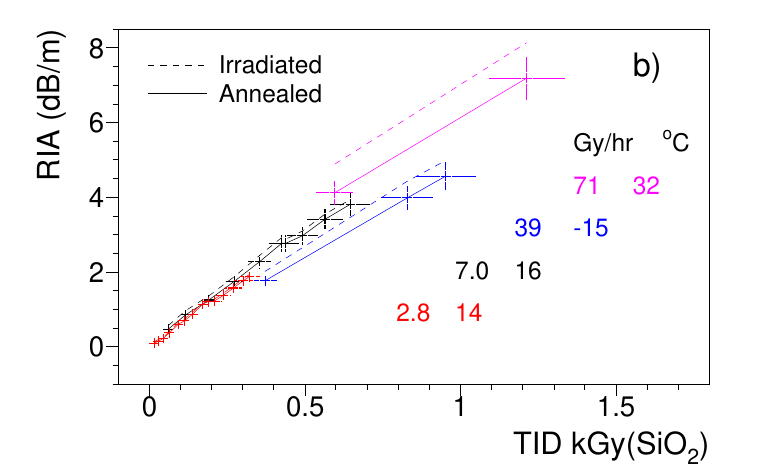}
    
    \caption{The RIA measurements versus TID
    are plotted for a) the Type-B OM2 MM, and
    b) Type-O OM4 MM fiber samples, at 
    dose-rates of 2.8 to 71~Gy(SiO$_2$)/hr and temperatures at -15 to 32~$^\circ$C.
    The dashed lines connect the RIAs versus the accumulated TIDs in consecutive days.
    The markers on solid-lines are the corresponding RIAs after 10 hours annealing. 
    The errors are estimated for 10~\% on the TIDs and 8~\% on the RIAs.
  \label{fig:compile-bsf_ofs} }
 
\end{figure}

The RIA measurements of consecutive days with
the two fiber types are 
compiled in Fig.~\ref{fig:compile-bsf_ofs}.a and b, respectively,
at the dose rates of 3 to 71~Gy(SiO$_2$)/hr
and the temperatures from -15 to 32~$^\circ$C.
The markers and lines of each color represent 
measurements of a fiber sample.
The dashed lines connect RIAs of samples at the highest TIDs of 
each day in irradiation.
The corresponding points on solid lines are 
the RIAs after 10 hours annealing.

In radiation, both fibers show similar characteristics
with the RIA reaching 4~dB/m at around 1~kGy(SiO$_2$).
At higher dose rates of $>40$ Gy/hr, the annealing recovery is 
at the 10 \% level. 
The measurements of samples are overlapping in two 
standard deviations of the means, that is,
the dose rate and temperature dependences are not significant
in the tested ranges.

\begin{figure}[b!] 
  \centering
    \includegraphics[width=.92\linewidth]
    {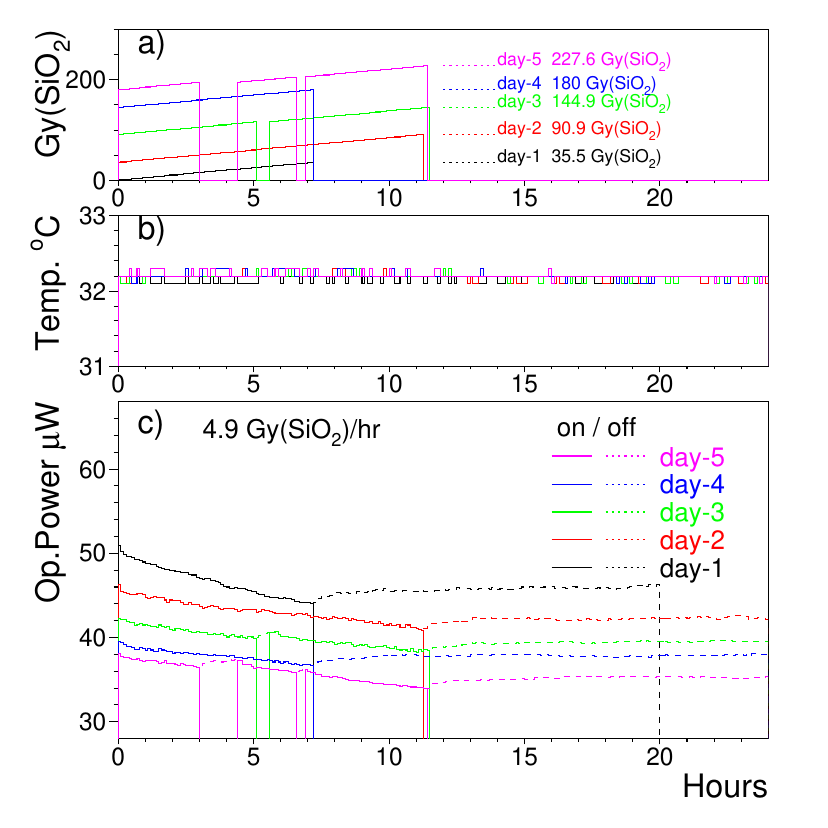}
    \vspace{-2mm}
    \caption{A Type-N Ge-doped OM3 MM fiber sample (400~m) 
    was irradiated in Co-60 gamma rays at a dose rate of
    4.9~Gy(SiO$_2$)/hr, at 32 $^\circ$C.
    The distributions of the initial five days are plotted for 
    a) the accumulated TIDs,
    b) the water tank temperatures, and c) the optical powers transmitted through
    the fiber sample in irradiation (solid lines) and 
    in annealing (dashed lines).
  \label{fig:cor5gy32oc} }

\vspace{4mm}  
  \centering
    \includegraphics[width=.92\linewidth]
    {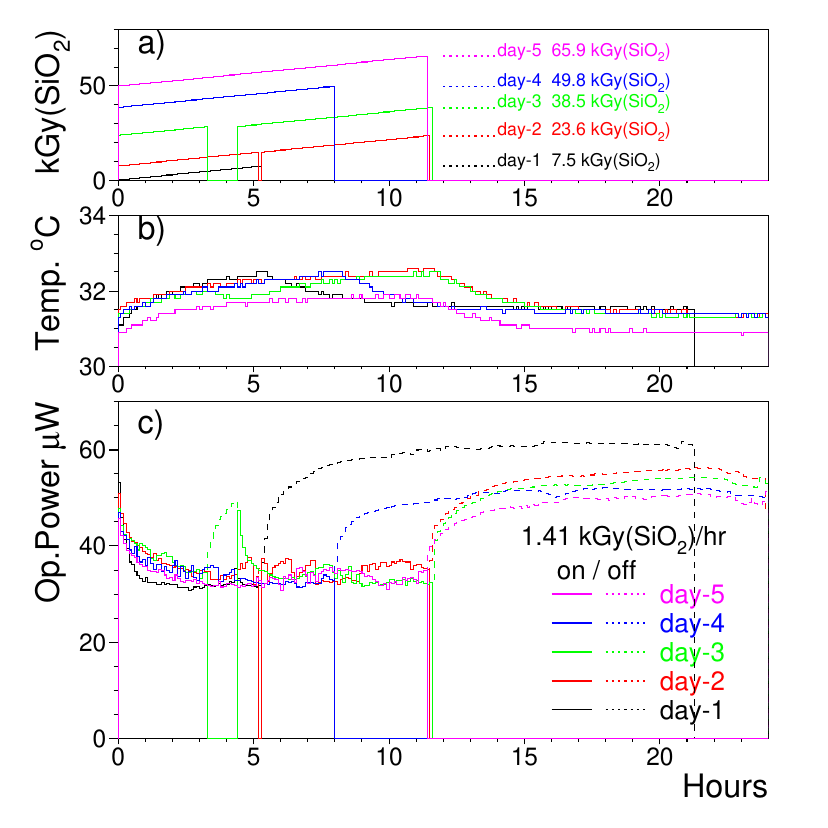}
    \vspace{-2mm}
    \caption{A type-M Ge-doped OM4 MM fiber sample (20~m) 
    was irradiated at a dose rate of
    1.41~kGy(SiO$_2$)/hr, at 31~$^\circ$C.
    The distributions recorded in the initial five days  
    are plotted for a) the accumulated TIDs, b)
    the sample temperature, and c) the optical power
    during irradiation (solid lines) and  
    in annealing (dashed lines).
  \label{fig:mmf1.41kgy31oc} }
\end{figure} 

\begin{figure}[b!] 
  \centering
    \includegraphics[width=.92\linewidth]{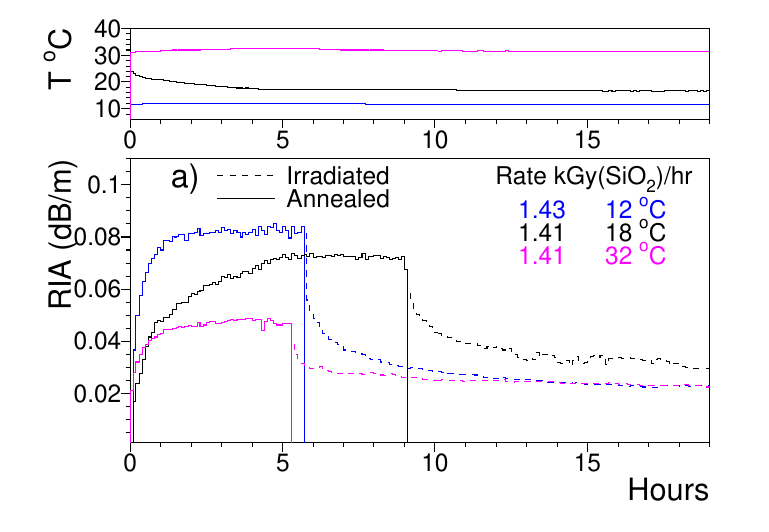} \\
    \vspace{-2mm}
    \includegraphics[width=.92\linewidth]{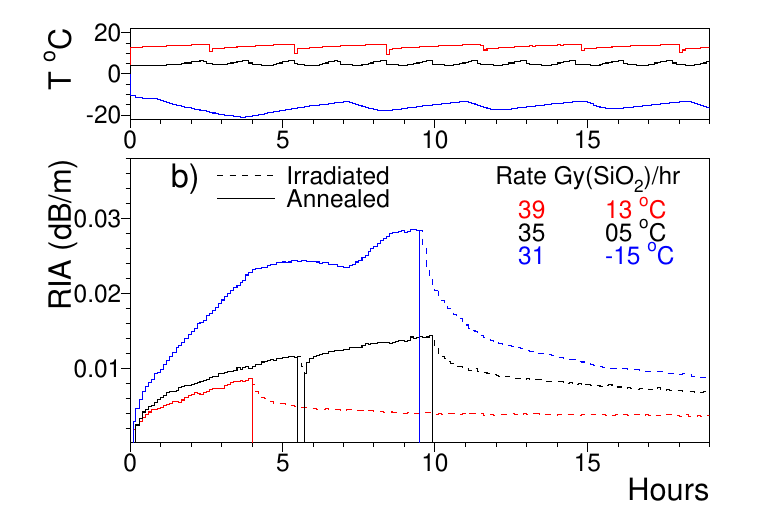}

    \caption{ Samples of Type-N OM3 MM fiber were irradiated in 
    Co-60 gamma rays. 
    The RIA distributions of the first days are plotted
    for the dose rates of a) $\sim$1.4~kGy/hr (OM3, 50~m), and
    b) $\sim$35~Gy/hr (OM3, 100~m). 
    The temperatures monitored are plotted on top of the RIA distributions.
    Each sample is indicated by a color.
    In a) the water tank was cooled
    by an external bath at constant temperatures.
    In b) the tank was chilled by a freezer plate; the 
    deviations on temperature due to the switch turning on/off are seen.
    The RIA distributions of the first day
    are plotted in solid lines during irradiation,
    and in annealing (dashed lines) after the source was shielded.
  \label{fig:corRatTmp} }
\end{figure} 

\begin{figure*}[t!] 
  \vspace{-10mm}      
  \centering
    \includegraphics[width=.98\linewidth]
    {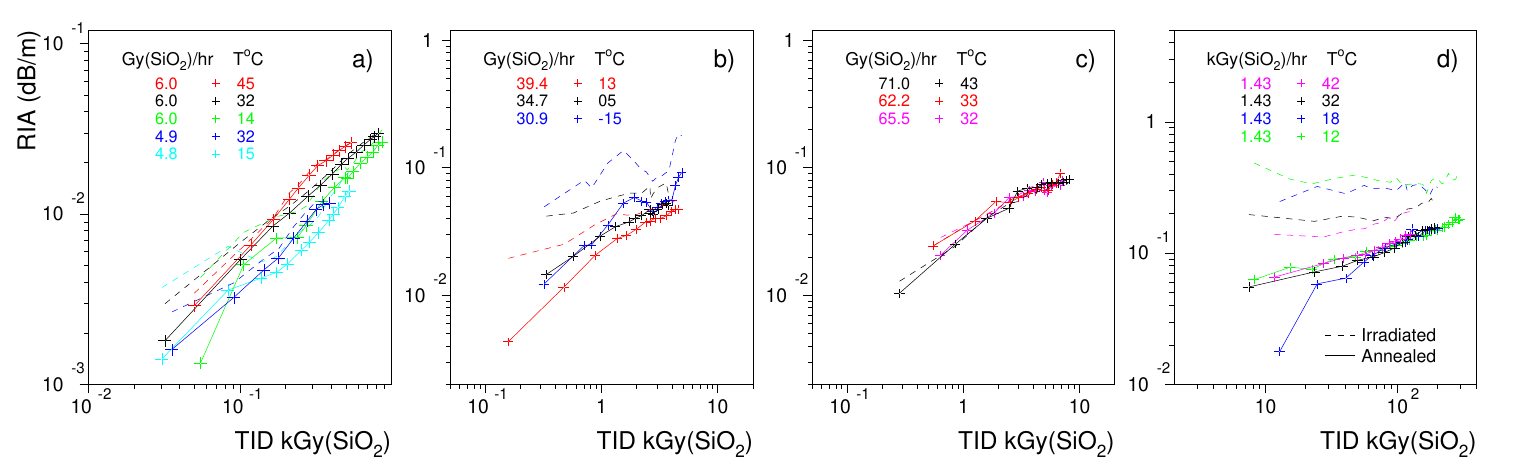} 
    \vspace{-2mm}
    \caption{Samples of Type-M Ge-doped OM3/OM4 MM fibers were irradiated in
    Co-60 gamma rays. 
    Plotted in a) to d) are the RIA distributions at dose rates of
    5 to 1.43k Gy(SiO$_2$)/hr, respectively, with the samples kept
    at  -15 to 45~$^\circ$C.
    Each of the dashed lines connects the highest instant 
    RIAs versus the accumulated TIDs
    of a sample in consecutive days.
    The markers on solid lines are the corresponding
    RIAs after 10 hours annealing.
  \label{fig:compile-mmf-tid} }
\vspace{2mm}
  \centering
    \includegraphics[width=.95\linewidth]
    {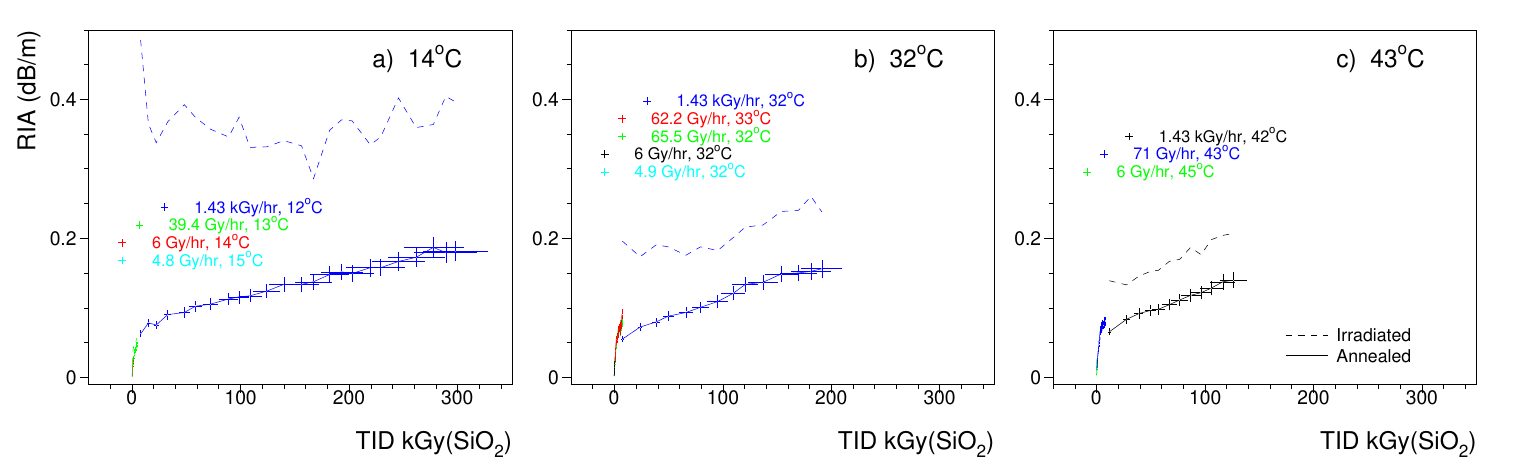} 
    \vspace{-2mm}
    \caption{The RIA distributions of Type-M OM3/OM4 MM fibers are compiled in
    a) to c) at temperatures of 14 to 43~$^\circ$C, respectively. 
    Each sample tested at a fixed dose rate and temperature  
    is presented with a dashed line connecting the highest instant RIAs 
    versus TIDs of consecutive days; the points on solid lines are the 
    corresponding RIAs after 10 hours annealing.
  \label{fig:compile-mmf-oc} }
\end{figure*} 

\begin{figure*}[t!] 
  \centering
    \includegraphics[width=.98\linewidth]
    {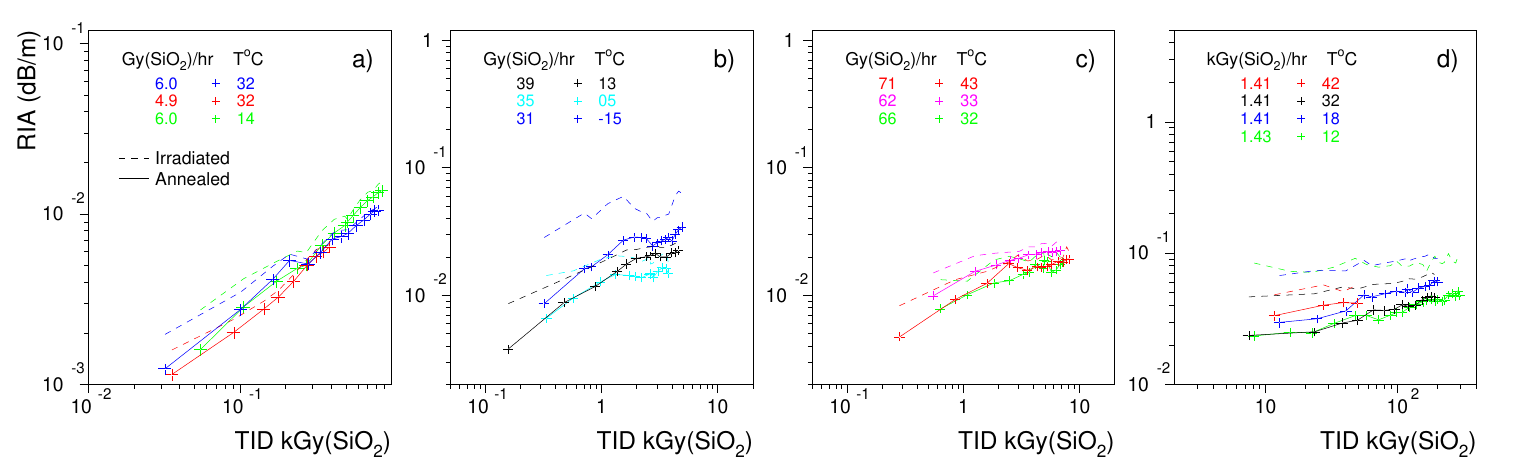} 
    \vspace{-2mm}
    \caption{Samples of Type-N Ge-doped OM3/OM4 MM fibers were irradiated in
    Co-60 gamma rays. 
    Plotted in a) to d) are the RIA distributions at dose rates of
    5 to 1.43k Gy(SiO$_2$)/hr, respectively, with the samples kept at 
    -15 to 45~$^\circ$C.
    Each of the dash lines connects the highest instant 
    RIAs versus the accumulated TIDs of 
    a sample in consecutive days.
    The points on solid lines are the corresponding RIAs after 10 hours 
    annealing.
  \label{fig:compile-corn-tid} }
\vspace{2mm}
  \centering
    \includegraphics[width=.95\linewidth]
    {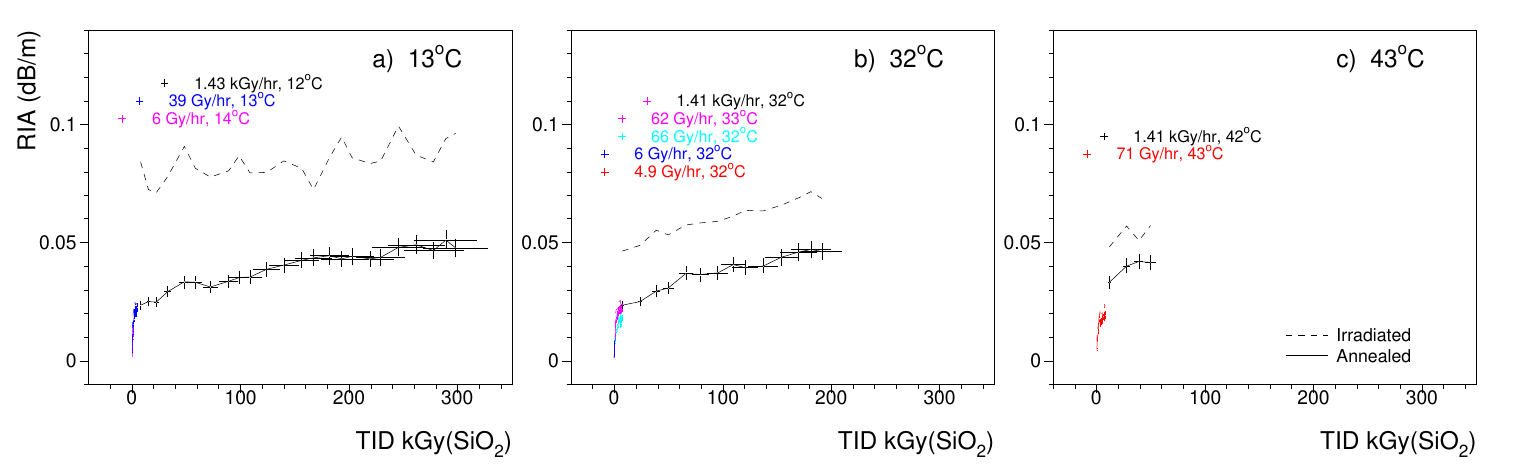}
    \vspace{-2mm}
    \caption{The RIA distributions of Type-N OM3/OM4 MM fibers are compiled in
    a) to c) at temperatures of 14 to 43~$^\circ$C, respectively. 
    Each sample  tested at a fixed dose rate and temperature  
    is presented with a dashed line connecting the 
    highest instant RIAs versus TIDs of consecutive days;
    the points on solid-lines are the corresponding RIAs
    after 10 hours annealing.
  \label{fig:compile-corn-oc} }
\end{figure*} 

\section{Radiation resistant fibers }  
\label{sec:Hard}

Two of the fiber types demonstrated radiation resistance 
on optical power attenuation  and recovery 
during and after irradiation, respectively.
The sensible dependences in radiation
were investigated for the dose rates of
5 to 1.43k~Gy(SiO$_2$)/hr, at temperatures from -15 to 45~$^\circ$C.

\subsection{Optical power measurements }  

At low dose rate, the radiation resistant fibers had
little optical power loss.
Long fibers were used to magnify the effects in irradiation.
Plotted in Fig.~\ref{fig:cor5gy32oc} are the measurements 
of a Type-N fiber (OM3, 400~m) at 4.9~Gy(SiO$_2$)/hr,
32~$^\circ$C.
The optical powers in the initial five days 
(Fig.~\ref{fig:cor5gy32oc}.c)
were decreasing linearly with the accumulated doses 
(Fig.~\ref{fig:cor5gy32oc}.a).
The annealing took about two hours, and
recovered by about 20 \% of the loss in each day.
This is indicating that the free ionizing defects generated are
floating for two hours before being recovered.

At high dose rate the ionizing defects are generated abundantly.
The measurements are shown for a 
Type-M (OM4, 20~m) fiber tested at 1.41~kGy(SiO$_2$)/hr, 31 $^\circ$C.
The optical powers in time of the initial five days are plotted
in Fig.~\ref{fig:mmf1.41kgy31oc}. 
The high dose rate had heated up the fiber sample by 2~$^\circ$C 
in the tank containing 0.5~liter of water 
(Fig.~\ref{fig:mmf1.41kgy31oc}.b).
The optical powers dropped quickly in the initial radiation 
(solid lines in Fig.~\ref{fig:mmf1.41kgy31oc}.c),
and the accumulated defects saturated in two hours
with the optical powers dropped to flat levels  (solid lines).
However, once the source was shielded, the annealing was fast 
and reached full recoveries in three hours (dashed lines).

\subsection{Dose rate and temperature dependences} 

The dose rate and temperature dependences of radiation 
resistant fibers were  evaluated with the RIA measurements
of the first irradiation days.
In Fig.~\ref{fig:corRatTmp}.a the RIAs in time
are plotted for samples of Type-N OM3 fiber 
at the high dose rates of $\sim$1.4~kGy(SiO$_2$)/hr and
the temperatures of 12 to 32~$^\circ$C, respectively.
The RIAs in radiation increased quickly in two hours to 
the saturation levels.
The RIA curve in black had a longer rising period, which was
caused by the temperature   being cooled 
from 24~$^\circ$C to 18~$^\circ$C

The temperature dependence is obvious. 
At 12~$^\circ$C the instant RIA level in radiation is 
twice higher than at 32~$^\circ$C. 
The annealing of both are effective to
compatible levels after 5 hours, which correspond to 
30\% (50\%) of the instant RIAs, respectively.

For comparison,
the measurements conducted at lower dose rates of $\sim$35~Gy(SiO$_2$)/hr 
in the first irradiation days
are plotted in Fig.~\ref{fig:corRatTmp}.b.
The radiation defects were accumulated slowly in time,
thus the RIAs increased continuously with  TIDs. 
The annealing had the RIAs recovered by about a third the magnitude
at warm temperatures.

At cold temperature, the stagnant annealing in radiation 
caused significant pileup of radiation defects.
The sample irradiated at \mbox{-15}~$^\circ$C
had twice higher instant RIA than those at warm temperatures.
However, once the source was shielded,
the annealing in cold was also effective with the RIA reduced
in half in a few hours.

The samples in Fig.~\ref{fig:corRatTmp}.b were chilled 
by a compressor. The relay switching on/off
caused glitches and large deviations in cold temperature.
For the setting of -15~$^\circ$C (blue line), 
the temperature was first chilled down to \mbox{-20}~$^\circ$C 
and was then heated up to -14~$^\circ$C, 
before being chilled again.   
The RIA curve (blue line) was effected by the warm-up 
and the annealing resulted to a dip on the RIA curve.


\subsection{RIA versus TID }   

The optical power attenuation of fibers has dependences
on dose rate and temperature.
The measurements of the two radiation resistant fibers
are compared for the RIAs as functions of TID. 

In Fig.~\ref{fig:compile-mmf-tid}.a to d, the RIA distributions 
of Type-M fibers are collected at
dose rates of 5 to 1.43k~Gy(SiO$_2$)/hr, respectively.
The measurements of each fiber sample is indicated by a color. 
The instant RIAs in radiation versus TIDs
are plotted in dashed lines, in consecutive days.  
The corresponding RIAs after 10 hours annealing 
are plotted in markers on solid lines.
The logarithmic scale helps to distinguish
data points from the slightly higher dashed lines
in low dose, warm temperature regions.

In Fig.~\ref{fig:compile-mmf-oc},
the RIA measurements are compiled for 
samples irradiated at near temperatures, 
in linear scale to demonstrated the logarithmic dependence on TID.
The RIA increases quickly at low total dose ($\sim$1~kGy),
and then turns to a slow slope to 0.2~dB/m at 300 kGy(SiO$_2$).
In Fig.~\ref{fig:compile-mmf-oc}.a,
the dashed line of the instant RIAs is twice higher 
than the annealed at 1.43~kGy/hr, 14~$^\circ$C.
For comparision, the instant RIA at the same dose rate, 32~$^\circ$C 
(dashed line in Fig.~\ref{fig:compile-mmf-oc}.b) 
has the magnitude reduced by half to the annealed.

The other type of radiation resistant fiber, the Type-N fiber,
has better RIA performance. The measurements  
complied at compatible dose rates and temperatures are plotted in
Fig.~\ref{fig:compile-corn-tid} and 
Fig.~\ref{fig:compile-corn-oc}, respectively.
The RIA also shows a quick
rise to 0.02~dB/m with the initial TID reaching $\sim$1~kGy,
and then a slow increase to 0.05~dB/m at 300~kGy(SiO$_2$).

The Type-N fiber could have  lower radiation defects generated, or 
better recovery efficiency.
The instant RIAs at high dose rate, 13~$^\circ$C,
adjoined by the dashed line in Fig.\ref{fig:compile-corn-oc}.a, 
are also a factor two higher than the annealed.

\section{Summary }
\label{sec:summary}

The telecom products of Ge-doped MM fibers provide 10 Gbps data
transmission for 850 nm fiber optics 
in distances of a few hundred meters.
The fiber fabrication methods matter for radiation hardness
required in nuclear applications.
With the four types of fibers tested, 
two of them show severe attenuation 
with the RIAs reaching 4~dB/m at the dose of around 1~kGy(SiO$_2$).

Two of the fiber types have shown radiation resistance and
dependences on dose rate and temperature,
with the annealed RIAs rising to 0.2 and 0.05~dB/m 
at the dose of 300~kGy(SiO$_2$), respectively.
The fiber samples were tested with OM3 and OM4 grades 
in different production batches. The batch dependence is not noticed.

The radiation induced defects  generated in fibers are
quickly recovered in a few hours.  
At low dose rate ($\sim$1~Gy/hr), the annealing effect is minor.
However, at high dose rate ($\sim$1~kGy/hr), 
the radiation induced defects are generated abundantly; 
the RIA in radiation is piled up twice higher than 
the annealed at room temperature. 
The recovery in cold is slower, which may increase the pile-up 
of defects to three times higher than the annealed.

The best performing type of fiber observed in this study
has the tolerance of 0.05~dB/m for the TID of 
a few hundreds kGy(SiO$_2$).
With the dose rate and temperature properly managed,
the applications in nuclear instrumentation is viable.

{\color{red}

}

\vspace{-2mm}
\section*{Acknowledgement}

The authors would like to thank the assistance of the Institute of Nuclear Energy Research.
This work has been partially supported by the Institute of Physics, Academia Sinica,
and the Ministry of Science and Technology, grant NSTC 113-2112-M-001-057. 

\vspace{-2mm}

{}


\begin{thebibliography}{}
\bibitem{OptoNIEL-2007} 
M.L. Chu, et al., 
``Radiation hardness studies of VCSELs and PINs for the 
opto-links of the Atlas SemiConductor Tracker'', 
Nucl. Instrum. Methods A 579 (2007) 795.

\bibitem{OptoNIEL-2011} 
S. Hou, et al., 
``Radiation hardness of optoelectronic components for the optical
readout of the ATLAS inner detector'', 
Nucl. Instrum. Methods A 636 (2011) S137.

\bibitem{CERN-2011}
J. Troska, et al.,
``Radiation Damage Studies of Lasers and Photodiodes
for Use in Multi-Gb/s Optical Data Links'',
IEEE TNS 58 (2011) 3103.

\bibitem{Johnston-2013}
A. H. Johnston,
``Radiation Effects in Optoelectronic Devices'',
IEEE TNS 60 (2013) 2054.

\bibitem{PD-2019}
L. Olantera, et al.,
``Radiation Effects on High-Speed InGaAs Photodiodes'',
IEEE TNS 66 (2019) 1663.


\bibitem{GBT} 
P. Moreira, et al., 
``The GBT Project'', 
in Proceedings, TWEPP 2009, DOI:10.5170/CERN-2009-006.342.

\bibitem{VTRX} 
J. Troska, et al., 
``Versatile Transceiver developments'', 
JINST 6 (2011) C01089.

\bibitem{VTRXplus} 
J. Troska, et al., 
``The VTRx+, an optical link module for data transmission at HL-LHC'', 
in Proceedings, TWEPP 2017, DOI:10.22323/1.313.0048.

 
\bibitem{King2010} 
M.P. King, et al.,  
``Response of a 0.25 $\mu$m thin-film silicon-on-sapphire CMOS technology to total
ionizing dose'',
JINST 5 (2010), C11021.

\bibitem{MTX-TID2024} 
D. Gong, et al., 
Characteristics of the MTx optical transmitter in Total Ionizing Dose, 
Nucl. Instrum. Methods A 1064 (2024) 169378.


\bibitem{MTX+} 
C.-P. Chao, et al., 
``Prototyping of a 25 Gbps optical transmitter for applications in high-energy physics experiments'', 
Nucl. Instrum. Methods A 979 (2020) 164399.

\bibitem{QTRX} 
B. Deng, et al., 
``A 40 Gbps optical transceiver for particle physics experiments'', 
JINST 17 (2022) C05005.

\bibitem{PAM4} 
L. Zhang, et al., 
``A 20 Gbps PAM4 data transmitter ASIC for particle physics experiments''
JINST 17 (2022) C03011.

\bibitem{GBT20} 
B. Deng, et al., 
``GBT20, a 20.48 Gbps PAM4 optical transmitter module
for particle physics experiments'', 
JINST 18 (2023) C02065.

\bibitem{Sensors2024}  
J. Li, et al., 
``Radiation Damage Mechanisms
and Research Status of RadiationResistant Optical Fibers: A Review'',
Sensors 2024, 24, 3235; DOI:10.3390/s24103235.

\bibitem{Girard2019}  
S. Girard, et al., 
``Overview of radiation induced point defects in 
silica-based optical fibers'',
Rev. Phys. 4 (2019) 100032.

\bibitem{Berghmans2007}  
F. Berghmans, et al., 
``An Introduction to Radiation 
Effects on Optical Components and Fiber Optic Sensors'',
in NATO Science for Peace and Security Series, 2007, DOI:10.1007/978-1-4020-6952-9\_6.

\bibitem{Oxford}  
D. Hall, et al., 
``The radiation induced attenuation of optical fibres
below -20 $^\circ$C exposed to lifetime HL-LHC doses at a
dose rate of 700 Gy(Si)/hr'',
JINST 7 (2012) C01047.


\bibitem{Ravotti} 
F. Ravotti, 
``Dosimetry Techniques and Radiation Test Facilities for Total Ionizing Dose Testing'',
IEEE Trans. Nucl. Sci. 65 (2018) 1440.

 



\end{thebibliography}
\end{document}